\pdfoutput=1

\newcommand*{\ATLASLATEXPATH}{}
\documentclass[cernpreprint,texlive=2016,txfonts,USenglish]{\ATLASLATEXPATH atlasdoc}

\usepackage{atlaspackage}

\usepackage{atlasbiblatex}

\usepackage{atlascontribute}

\usepackage{atlasphysics}

\usepackage{siunitx}
\usepackage[utf8]{inputenc}
\usepackage[T1]{fontenc} 

\usepackage{defs}

\AtlasTitle{Evidence for light-by-light scattering in heavy-ion collisions with the ATLAS detector at the LHC}

\author{The ATLAS Collaboration}

\PreprintIdNumber{CERN-EP-2016-316}

\AtlasJournal{Nature Physics}
 \AtlasJournalRef{Nature Phys. 13 (2017) no.9, 852-858}
 \AtlasDOI{10.1038/nphys4208}

\AtlasAbstract{
Light-by-light scattering ($\gamma\gamma\rightarrow\gamma\gamma$) is a quantum-mechanical process that is forbidden in the classical theory of electrodynamics.
This reaction is accessible at the Large Hadron Collider thanks to the large electromagnetic field strengths generated by ultra-relativistic colliding lead ions.
Using 480~$\mu$b$^{-1}$~of lead--lead collision data recorded at a centre-of-mass energy per nucleon pair of $5.02$~\TeV{} by the ATLAS detector, here we report evidence for light-by-light scattering.
A total of thirteen candidate events were observed with an expected background of $2.6\pm0.7$ events. 
After background subtraction and analysis corrections, the fiducial cross section of the process 
$\textrm{Pb+Pb}\,(\gamma\gamma)\rightarrow \textrm{Pb}^{(\ast)}\textrm{+}\textrm{Pb}^{(\ast)}\,\gamma\gamma$,
for photon transverse energy $E_{\mathrm{T}}>3~\gev$, photon absolute pseudorapidity $|\eta|<2.4$, diphoton invariant mass greater than $6~\gev$, 
diphoton transverse momentum lower than 2~\GeV{} and diphoton acoplanarity below 0.01, is measured to be
$70 \pm 24~\textrm{(stat.)} \pm 17~\textrm{(syst.)}$ nb, which is in agreement with the Standard Model predictions.
}

\hypersetup{pdftitle={ATLAS document},pdfauthor={The ATLAS Collaboration}}

\begin{document}

\maketitle

\begin{figure}[b!]
\begin{center}
	\includegraphics[width=1.0\textwidth]{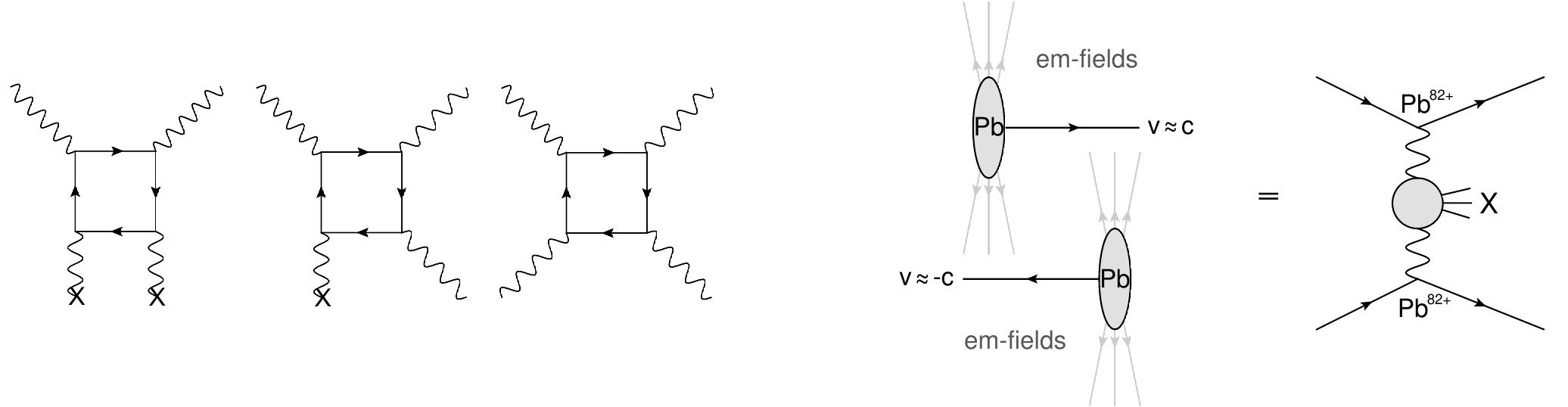}\put(-355,-8){{\footnotesize(a)}}\put(-93,-8){{\footnotesize(b)}}
	\caption{Diagrams illustrating the QED LbyL interaction processes and the equivalent photon approximation. (a) Diagrams for Delbrück scattering (left), photon splitting (middle) and elastic LbyL scattering (right). 
	Each cross denotes external field legs, e.g., an atomic Coulomb field or a strong background magnetic field. 
	(b) Illustration of an ultra-peripheral collision of two lead ions.
	Electromagnetic interaction between the ions can be described as an exchange of photons that can couple to form a given final-state X. 
	The flux of photons is determined from the Fourier transform of the electromagnetic field of the ion, taking into account the nuclear electromagnetic form factors.}
	\label{fig:feyncorrections}
\end{center}
\end{figure}

One of the key features of Maxwell's equations is their linearity in both the sources and the fields, from which follows the superposition principle.
This forbids effects such as light-by-light (LbyL) scattering, $\gamma\gamma\rightarrow\gamma\gamma$, which is a purely quantum-mechanical process.
It was realised in the early history of quantum electrodynamics (QED) that LbyL scattering is related to the polarisation of the vacuum~\cite{Heisenberg1936}.
In the Standard Model (SM) of particle physics, the virtual particles that mediate the LbyL coupling are electrically charged fermions 
or $W^{\pm}$ bosons.
In QED, the $\gamma\gamma\rightarrow \gamma\gamma$ reaction proceeds at lowest order in the fine structure constant ($\alpha_{\textrm{em}}$) via virtual one-loop box diagrams involving fermions (Figure~\ref{fig:feyncorrections}(a)), which is an $\mathcal{O}(\alpha_{\textrm{em}}^4\approx3\times 10^{-9})$
process, making it challenging to test experimentally.
Indeed, the elastic LbyL scattering has remained unobserved: even the ultra-intense laser experiments are not yet powerful enough to probe this phenomenon~\cite{King:2015tba}.

LbyL scattering via an electron loop has been precisely, albeit indirectly, tested in measurements of the anomalous magnetic moment of the electron and muon~\cite{Hanneke:2008tm, Bennett:2006fi} where it is predicted to contribute substantially, as one of the QED corrections~\cite{Laporta:1992pa}. 
The $\gamma\gamma\rightarrow \gamma\gamma$ reaction has been measured in photon scattering in the Coulomb field of a nucleus (Delbrück scattering) at fixed photon energies below 7~\GeV~\cite{Wilson:1953zz,PhysRevD.8.3813,Schumacher:1975kv,PhysRevC.58.2844}.
The analogous process, where a photon splits into two photons by interaction with external fields (photon splitting), has been observed in the energy region of  0.1--0.5~\GeV~\cite{Akhmadaliev:2001ik}.
A related process involving only real photons, in which several photons fuse to form an electron--positron pair ($e^+e^-$), has been measured in Ref.~\cite{Burke:1997ew}. 
Similarly, the multiphoton Compton scattering in which up to four laser photons interact with an electron, has been observed~\cite{Bula:1996st}.

An alternative way by which LbyL interactions can be studied is by using relativistic heavy-ion collisions.  
In `ultra-peripheral collision' (UPC) events, with impact parameters larger than twice the radius of the nuclei~\cite{Bertulani:1987tz, Natale:2001zj}, the strong interaction does not play a role.  
The electromagnetic (EM) field strengths of relativistic ions scale with the proton number ($Z$). For example, for a lead nucleus (Pb) with $Z=82$ the field can be up to $10^{25}\,\textrm{Vm}^{-1}$~\cite{Sengul:2015ira}, much larger than the Schwinger limit~\cite{Schwinger:1951nm} above which QED corrections become important.
In the 1930s it was found that highly relativistic charged particles can be described by the equivalent photon approximation (EPA)~\cite{Fermi:1925fq,vonWeizsacker:1934nji,PhysRev.45.729}, which is schematically shown in Figure~\ref{fig:feyncorrections}(b).
The EM fields produced by the colliding Pb nuclei 
can be treated as a beam of quasi-real photons with a small virtuality of $Q^2 < 1/R^2$, where $R$ is the radius of the charge distribution and so $Q^2 < 10^{-3}~\GeV^2$.
Then, the cross section for the reaction $\textrm{Pb+Pb}\,(\gamma\gamma)\rightarrow \textrm{Pb}\textrm{+}\textrm{Pb}\,\gamma\gamma$ can be calculated by convolving the respective photon flux with the elementary cross section for the process $\gamma\gamma\rightarrow \gamma\gamma$.
Since the photon flux associated with each nucleus scales as $Z^2$, the cross section is extremely enhanced as compared to proton--proton ($pp$) collisions.

In this article a measurement of LbyL scattering in Pb+Pb collisions at the Large Hadron Collider (LHC) is reported, following the approach recently proposed in Ref.~\cite{Enterria:2013yra}.
The final-state signature of interest is the exclusive production of two photons, 
$\textrm{Pb+Pb}\,(\gamma\gamma)\rightarrow \textrm{Pb}^{(\ast)}\textrm{+}\textrm{Pb}^{(\ast)}\,\gamma\gamma$, where a possible EM excitation of the outgoing ions~\cite{ALICE:2012aa} is denoted by $(^{\ast})$. 
Hence, the expected signature is two
photons 
and no further activity in the central detector, since the $\textrm{Pb}^{(\ast)}$ ions escape into the LHC beam pipe.
Moreover, it is predicted that the background is relatively low in
heavy-ion collisions and is dominated by exclusive dielectron (\ggee) production~\cite{Enterria:2013yra,Knapen:2016moh}. The misidentification of electrons as photons can occur when the electron track is not reconstructed or the electron emits a hard bremsstrahlung photon.
The fiducial 
cross section of the process $\gamma\gamma\rightarrow \gamma\gamma$ in $\textrm{Pb+Pb}$ collisions is measured, using a data set recorded at a nucleon--nucleon centre-of-mass energy (\sqrtNN) of $5.02$~\TeV. 
This data set was recorded with the ATLAS detector at the LHC in 2015 and corresponds to an integrated luminosity of $480\pm 30~\invmub$.
In addition to the measured fiducial cross section, the significance of the observed number of signal candidate events is given, assuming the background-only hypothesis.

\section*{\label{sec:Atlas}Experimental setup}

ATLAS is a cylindrical particle detector composed of several sub-detectors~\cite{ATLAS_detector_paper}. ATLAS uses a right-handed coordinate system with its origin at the nominal interaction point (IP)
in the centre of the detector and the $z$-axis along the beam pipe.
The $x$-axis points from the IP to the centre of the LHC ring,
and the $y$-axis points upwards.
Cylindrical coordinates $(r,\phi)$ are used in the transverse plane, 
$\phi$ being the azimuthal angle around the $z$-axis.
The pseudorapidity is defined in terms of the polar angle $\theta$ as $\eta = -\ln \tan(\theta/2)$. Angular distance is measured in units of $\Delta R \equiv \sqrt{(\Delta\eta)^{2} + (\Delta\phi)^{2}}$.
The photon or electron transverse energy is $\et= E \sin(\theta)$, where $E$ is its energy. 
The inner tracking detector (ID) consists of a silicon pixel system, a silicon microstrip detector and a straw-tube tracker immersed in a 2\,T magnetic field provided by a superconducting solenoid. 
The ID track reconstruction efficiency is estimated in Ref.~\cite{Aaboud:2016itf} for minimum-bias $pp$ events that, like UPC Pb+Pb events, have a low average track multiplicity. 
For charged hadrons in the transverse momentum range $100<\pt<200$~\MeV~the efficiency is about 50\% and grows to 80\% for $\pt>200$~\MeV.
Around the tracker there is a system of EM and hadronic calorimeters, which use liquid argon and lead, copper, or tungsten absorbers for the EM and forward ($|\eta| > 1.7$) hadronic components of the detector, and scintillator-tile active material and steel absorbers for the central ($|\eta|< 1.7$) hadronic component.
The muon spectrometer (MS) consists of separate trigger and
high-precision tracking chambers measuring the trajectory of muons in a magnetic field generated by superconducting air-core toroids.
The ATLAS minimum-bias trigger scintillators (MBTS) consist of scintillator slabs positioned between the ID and the endcap calorimeters with each side having an outer ring of four slabs segmented in azimuthal angle, covering $2.07 < |\eta| < 2.76$ and
an inner ring of eight slabs, covering $2.76 < |\eta| < 3.86$.
The ATLAS zero-degree calorimeters (ZDC), located along the beam axis at 140 m from the IP on both sides, detect neutral particles (including neutrons emitted from the nucleus).  
The ATLAS trigger system~\cite{Aad:1388600} consists of a Level-1 trigger implemented using a combination of dedicated electronics and programmable logic, and a software-based high-level trigger (HLT).

\section*{\label{sec:DataMC}Monte Carlo simulation and theoretical predictions}

Several Monte Carlo (MC) 
samples are produced to estimate background contributions and corrections to the fiducial measurement.
The detector response is modelled using a simulation based on GEANT4~\cite{Agostinelli:2002hh,Aad:2010ah}.
The data and MC simulated events are passed through the same reconstruction and analysis procedures.

Light-by-light signal events are generated taking into account box diagrams with charged leptons and quarks in the loops, as detailed in Ref.~\cite{Klusek-Gawenda:2016euz}.
The contributions from $W\textrm{-boson}$ loops are omitted in the calculations since they are mostly important for diphoton masses $m_{\gamma\gamma} > 2m_{W}$~\cite{Fichet:2014uka}.
The calculations are then convolved with the \PbPb\ EPA spectrum from the \starlight 1.1 MC generator~\cite{Klein:2016yzr}.
Next, various diphoton kinematic distributions are cross-checked with predictions from Ref.~\cite{Enterria:2013yra} and good agreement is found.
The theoretical uncertainty on the cross section is mainly due to limited knowledge of the nuclear electromagnetic form-factors 
and the related initial photon fluxes. This is studied in Ref.~\cite{Enterria:2013yra} and the relevant uncertainty is conservatively estimated to be $20\%$.
Higher-order corrections (not included in the calculations) are also part of the theoretical uncertainty and are of the order of a few percent for diphoton invariant masses below 100~\GeV~\cite{Bern:2001dg, Klusek-Gawenda:2016nuo}.

The sources of background considered in this analysis are:
\ggee, central exclusive production (CEP) of photon pairs, exclusive production of quark--antiquark pairs ($\gamma\gamma\rightarrow q\bar{q}$) and other backgrounds that could mimic the diphoton event signatures.
The \ggee{} background is modelled with \starlight 1.1~\cite{Klein:2016yzr}, 
in which the  cross section is computed by combining the \PbPb\ EPA with the leading-order formula for \ggee.
This process has been recently measured by the ALICE Collaboration, and a good agreement with \starlight is found~\cite{Abbas:2013oua}.
The exclusive diphoton final state can be also produced via the strong interaction through a quark loop in the exchange of two gluons in a colour-singlet state.
This CEP process, $gg\rightarrow\gamma\gamma$, is modelled using \superchic 2.03~\cite{Harland-Lang:2015cta}, in which the $pp$ cross section has been scaled by $A^2R_g^4$ as suggested in Ref.~\cite{Enterria:2013yra}, where $A=208$ and $R_g\approx0.7$ is a gluon shadowing correction~\cite{Eskola:2009uj}.
This process has a large theoretical uncertainty, of $\mathcal{O}(100\%)$, mostly related to incomplete knowledge of gluon densities~\cite{Aaltonen:2011hi}. 
The $\gamma\gamma\rightarrow q\bar{q}$ contribution is estimated using \textsc{Herwig}++ 2.7.1~\cite{Bahr:2008pv} where the EPA formalism in $pp$ collisions is implemented. 
The $\gamma\gamma\rightarrow q\bar{q}$ sample is then normalised to the corresponding cross section in \PbPb\ collisions~\cite{Klein:2016yzr}.

\section*{\label{sec:Selection}Event selection}

Candidate diphoton events were recorded in the \PbPb\ run in 2015 using a dedicated trigger for events with moderate activity in the calorimeter but little additional activity in the entire detector. 
At Level-1 the total \et\ registered in the calorimeter after noise suppression was required to be between $5$ and $200~\gev$. 
Then at the HLT, events were rejected if more than one hit was found in the inner ring of the MBTS (MBTS veto) or if more than ten hits were found in the pixel detector.

The efficiency of the Level-1 trigger is estimated with \ggee\ events passing an independent supporting trigger.
This trigger is designed to select events with mutual dissociation of Pb nuclei and small activity in the ID. 
It is based on a coincidence of signals in both ZDC sides and a requirement on the total \et\ in the calorimeter below 50~\GeV.
Event candidates are required to have only two reconstructed tracks and two EM energy clusters.
Furthermore, to reduce possible backgrounds, each pair of clusters (cl1, cl2) is required to have a small acoplanarity ($1-\Delta\phi_{\textrm{cl1,~cl2}}/\pi<0.2$).
The extracted Level-1 trigger efficiency is provided as a function of the sum of cluster transverse energies (\etclone+\etcltwo). 
The efficiency grows from about $70\%$ at $(\etclone+\etcltwo)=6~\gev$ to $100\%$ at $(\etclone+\etcltwo)>9~\gev$.
The efficiency is parameterised using an error function fit which is then used to reweight the simulation.
Due to the extremely low noise, very high hit reconstruction efficiency and low conversion probability of signal photons in the pixel detector (around 10\%), the uncertainty due to the requirement for minimal activity in the ID is negligible.
The MBTS veto efficiency was studied using \ggll~events ($\ell=e,~\mu$) passing a supporting trigger and it is estimated to be $(98\pm2)\%$.

Photons are reconstructed from EM clusters in the calorimeter and tracking information provided by the ID, which allows the identification of photon conversions. 
Selection requirements are applied to remove EM clusters with a large amount of energy from poorly functioning calorimeter cells, and a timing requirement is made to reject out-of-time candidates. An energy calibration specifically optimised for photons~\cite{Aad:2014nim} is applied to the candidates to account for upstream energy loss and both lateral and longitudinal shower leakage. 
A dedicated correction~\cite{Aaboud:2016yuq} is applied for photons in MC samples to correct for potential mismodelling of quantities which describe the properties (``shapes'') of the associated EM showers.

The photon particle-identification~(PID) in this analysis is based on three shower-shape variables: the lateral width of the 
shower in the middle layer of the EM calorimeter, the ratio of the energy difference associated with the largest and second 
largest energy deposits to the sum of these energies in the first layer, and the fraction of energy reconstructed in the first layer relative to the total energy of the cluster. 
Only photons with $\et>3~\gev$ and $|\eta|<2.37$, excluding the calorimeter transition region $1.37<|\eta|<1.52$, are considered. The pseudorapidity requirement ensures that the photon candidates pass through regions of the EM calorimeter where the first layer is segmented into narrow strips, allowing for good separation between genuine prompt photons and photons coming from the decay of neutral hadrons. 
A constant photon PID efficiency of 95\% as a function of $\eta$ with respect to reconstructed photon candidates is maintained. 
This is optimised using multivariate analysis techniques~\cite{Hocker:2007ht}, such that EM energy clusters induced by cosmic-ray muons are rejected with 95\% efficiency.

Preselected events are required to have exactly two photons satisfying the above selection criteria, with a diphoton invariant mass greater than 6~\GeV.
In order to reduce the dielectron background, a veto on the presence of any charged-particle tracks (with $\pt>100$~\MeV{}, $|\eta|<2.5$ and at least one hit in the pixel detector) is imposed. 
This requirement further reduces the fake-photon background from the dielectron final state by a factor of 25, according to simulation.
It has almost no impact on \gggg\ signal events, since the probability of photon conversion in the pixel detector is relatively small and converted photons are suppressed at low \et\ (3--6~\GeV) by the photon selection requirements.
According to MC studies, the photon selection requirements remove about 10\% of low-\et\ photons.
To reduce other fake-photon backgrounds (for example, cosmic-ray muons), the transverse momentum of the diphoton system~(\ptgg) is required to be below 2~\GeV. 
To reduce background from CEP $gg\rightarrow\gamma\gamma$ reactions, an additional requirement on diphoton acoplanarity, $\textrm{Aco}=1-\Delta\phi_{\gamma\gamma}/\pi<0.01$, is imposed.
This requirement is optimised to retain a high signal efficiency and reduce the CEP background significantly, since the transverse momentum transferred by the photon exchange is usually much smaller than that due to the colour-singlet-state gluons~\cite{Baur:2001jj}.

\section*{\label{sec:DetCalib}Performance and validation of photon reconstruction}

Since the analysis requires the presence of low-energy 
photons, which are not typically used in ATLAS analyses,
detailed studies of photon reconstruction and calibration are performed.

\begin{figure}[htb]
\begin{center}
\includegraphics[width=0.85\textwidth]{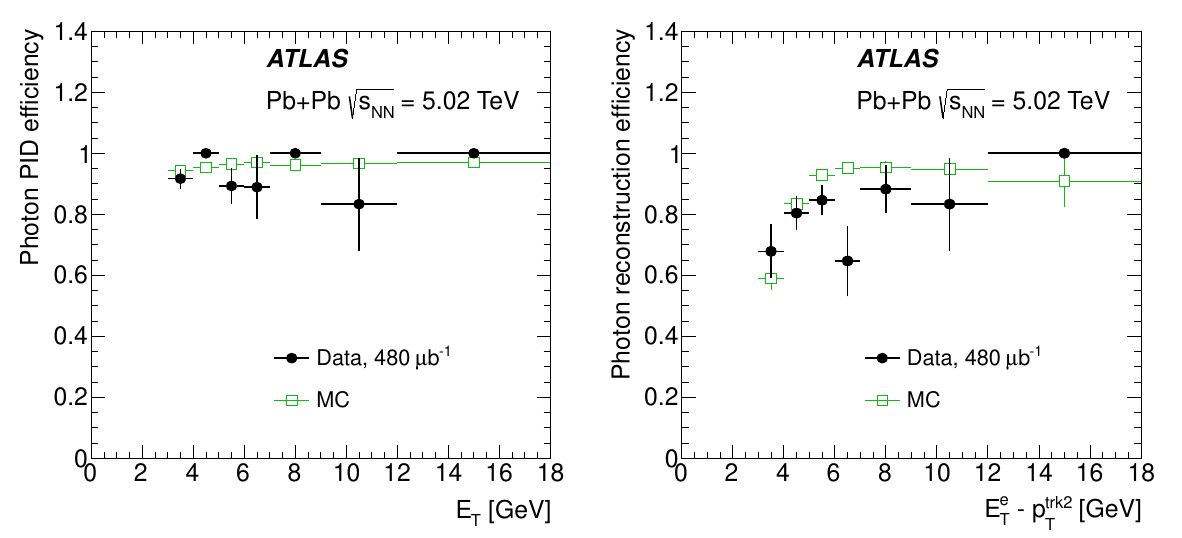}\put(-295,-8){{\footnotesize(a)}}\put(-95,-8){{\footnotesize(b)}}
\caption{Photon identification and reconstruction efficiencies. (a) Photon PID efficiency as a function of photon \et\ extracted from FSR event candidates. (b) Photon reconstruction efficiency as a function of photon \eT\ (approximated with $\eT^e - \pT^\textrm{trk2}$) extracted from \ggee events with a hard-bremsstrahlung photon. 
Data (closed markers) are compared with MC simulations (open markers).
The statistical uncertainties arising from the finite size of the data and simulation samples are indicated by vertical bars.}
\label{fig:pt-fsr-eff}
\end{center}
\end{figure}

High-\pt\ \ggll production with a final-state radiation~(FSR) photon is used for the measurement of the photon PID 
efficiency. Events with a photon and two tracks corresponding to oppositely charged particles with $\pt>1~\gev$ are required to pass the same trigger as in the diphoton selection or the supporting trigger.
The $\Delta R$ between a photon candidate and a track is required to be greater than 0.2 in order to avoid leakage of the electron clusters from the \ggee\ process to the photon cluster. 
The FSR event candidates are identified using a $\pTttg<1~\gev$ requirement, where \pTttg\ is the transverse momentum of the three-body system consisting of two charged-particle tracks and a photon. 
The FSR photons are then used to extract the photon PID efficiency, which is defined as the probability for a reconstructed photon to satisfy the identification criteria. 
Figure~\ref{fig:pt-fsr-eff}(a) shows the photon PID efficiencies in data and simulation as a function of reconstructed photon \et. 
Within their statistical precision the two results agree.

The photon reconstruction efficiency is extracted from data using \ggee{} events where one of the electrons emits a hard-bremsstrahlung photon due to interaction with the material of the detector.
Events with exactly one identified electron, two reconstructed charged-particle tracks and exactly one photon are studied.
The electron \eT is required to be above 5~\GeV{} and the \pT of the track that is unmatched with the electron (trk2) is required to be below $2$~\GeV{}.
The additional hard-bremsstrahlung photon is expected to have $\eT^{\gamma}\approx (\eT^e-\pT^\textrm{trk2})$.
The $\pT^\textrm{trk2}<2$~\GeV{} requirement ensures a sufficient $\Delta R$ separation between the expected photon and the second electron, extrapolated to the first layer of the EM calorimeter.
The data sample contains 247 $\ggee$ events that are used to extract the photon reconstruction efficiency, which is presented in Figure~\ref{fig:pt-fsr-eff}(b).
Good agreement between data and \ggee~MC simulation is observed and the photon reconstruction efficiency is measured with a 5--10\% relative uncertainty at low \eT (3--6~\GeV).

In addition, a cross-check is performed on $Z\rightarrow\mu^+\mu^-\gamma$ events identified in \pp\ collision data from 2015 corresponding to an integrated luminosity of $1.6\,\text{fb}^{-1}$. 
The results support (in a similar way to Ref.~\cite{ATLAS-CONF-2012-143}) the choice to use the three shower-shape variables in this photon PID selection in an independent sample of low-\et photons.

The photon cluster energy resolution is extracted from data using \ggee\ events.
The electrons from the \ggee\ reaction are well balanced in their transverse momenta, with very small standard deviation, $\sigma_{\pt^{e+}-\pt^{e-}}<30 $~\MeV{}, much smaller than the expected EM calorimeter energy resolution.
Therefore, by measuring ($\etclone-\etcltwo$) distributions in \ggee\ events, one can extract the cluster energy resolution, $\sigma_{\et^\textrm{cl}}$.
For electrons with $\et <10$~\GeV{} the $\sigma_{\et^\textrm{cl}}/\et^\textrm{cl}$ is observed to be approximately $8\%$ both in data and simulation. 
An uncertainty of $\delta \sigma_{\et^{\gamma}}/\sigma_{\et^{\gamma}} = 15\%$ is assigned to the simulated photon energy resolution and takes into account differences between $\sigma_{\et^\textrm{cl}}$ in data and $\sigma_{\et^{\gamma}}$ in simulation.

Similarly, the EM cluster energy scale can be studied using the ($\etclone+\etcltwo$) distribution. 
It is observed that the simulation provides a good description of this distribution, within the relative uncertainty of 5\% that is assigned to the EM cluster energy-scale modelling.

\section*{\label{sec:Background}Background estimation}

Due to its relatively high rate, the exclusive production of electron pairs ($\ggee$) can be a source of fake diphoton events.
The contribution from the dielectron background is estimated using \ggee\ MC simulation (which gives 1.3 events) and is verified using the following data-driven technique.
Two control regions are defined that are expected to be dominated by \ggee\ backgrounds.
The first control region is defined by requiring events with exactly one reconstructed charged-particle track 
and two identified photons that satisfy the same preselection criteria as for the signal definition. 
The second control region is defined similarly to the first one, except exactly two tracks are required ($\Ntrk=2$).
Good agreement is observed between data and MC simulation in both control regions, but the precision is limited by the number of events in data. A conservative uncertainty of 25\% is therefore assigned to the \ggee background estimation, which reflects the statistical uncertainty of data in the $\Ntrk=1$ control region. 
The contribution from a related QED process, $\gamma\gamma\rightarrow e^+e^-\gamma\gamma$, is evaluated using the \textsc{MadGraph5$\_$aMC$@$NLO} MC generator~\cite{Alwall:2014hca} and is found to be negligible.

The Aco $<0.01$ requirement significantly reduces the CEP $gg\rightarrow\gamma\gamma$ background. 
However, the MC prediction for this process has a large theoretical uncertainty, hence an additional data-driven normalisation is performed in the region Aco $>b$, where $b$ is a value greater than 0.01 which can be varied.
Three values of $b$ (0.01,~0.02,~0.03) are used, where the central value $b=0.02$ is chosen to derive the nominal background prediction and the values $b=0.01$ and $b=0.03$ to define the systematic uncertainty. 
The normalisation is performed using the condition:
$f_{gg\rightarrow\gamma\gamma}^{\textrm{norm}, b} = \bigl( N_\textrm{data}(\textrm{Aco} > b) - N_\textrm{sig}(\textrm{Aco} > b) - N_{\gamma\gamma\rightarrow e^+e^-}(\textrm{Aco} > b)\bigr) / N_{gg\rightarrow\gamma\gamma}(\textrm{Aco} > b)$,
for each value of $b$, where $N_\textrm{data}$  is the number of observed events, $N_\textrm{sig}$ is the expected number of signal events, 
$N_{\gamma\gamma\rightarrow e^+e^-}$ is the expected background from $\gamma\gamma\rightarrow e^+e^-$ events and $N_{gg\rightarrow\gamma\gamma}$ is the MC estimate of the expected background from CEP $gg\rightarrow\gamma\gamma$ events.
The normalisation factor is found to be $f_{gg\rightarrow\gamma\gamma}^{\textrm{norm}}= 0.5\pm 0.3$
and the background due to CEP $gg\rightarrow\gamma\gamma$ is estimated to be $f_{gg\rightarrow\gamma\gamma}^{\textrm{norm}} \times N_{gg\rightarrow\gamma\gamma}(\textrm{Aco} <0.01) = 0.9 \pm 0.5$ events. 
To verify the CEP $gg\rightarrow\gamma\gamma$ background estimation method, energy deposits in the ZDC are studied for events before the Aco selection. 
It is expected that the outgoing ions in CEP events predominantly dissociate, which results in the emission of neutrons detectable in the ZDC~\cite{Enterria:2013yra}.
Good agreement between the normalised CEP $gg\rightarrow\gamma\gamma$ MC expectation and the observed events with a ZDC signal corresponding to at least 1 neutron is observed in the full Aco range.

Low-\pt dijet events can produce multiple $\pi^0$ mesons which could potentially mimic diphoton events.
The event selection requirements are efficient in rejecting such events, and based on studies performed with a supporting trigger, 
the background from hadronic processes is estimated to be $0.3\pm0.3$ events. 
MC studies show the background from \ggqq\ processes is negligible.

Exclusive neutral two-meson production can be a potential source of background for LbyL events, mainly due to their back-to-back topology being similar to that of the CEP $gg\rightarrow\gamma\gamma$ process.
The cross section for this process is calculated to be below 10\% of the CEP $gg\rightarrow\gamma\gamma$ cross section~\cite{HarlandLang:2011qd,Harland-Lang:2013ncy} and it is therefore considered to give a negligible contribution to the signal region.
The contribution from bottomonia production (for example, $\gamma\gamma\rightarrow \eta_b \rightarrow \gamma\gamma$ or $\gamma\textrm{Pb}\rightarrow \Upsilon \rightarrow \gamma\eta_b \rightarrow 3\gamma$) is calculated using parameters from Refs.~\cite{Ebert:2002pp, Segovia:2016xqb} and is found to be negligible.

The contribution from other fake diphoton events (for example those induced by cosmic-ray muons) is estimated using 
photons that fail to satisfy the longitudinal shower-shape requirement.
The total background due to other fake photons is found to be $0.1 \pm 0.1$~events.
As a further cross-check, additional activity in the MS is studied. It is observed that out of 18 events satisfying the inverted $\pt^{\gamma\gamma}$ requirement, 13 have at least one additional reconstructed muon.
In the region $\pt^{\gamma\gamma} < 2~\GeV$, no events with muon activity are found, which is compatible with the above-mentioned estimate of $0.1 \pm 0.1$.

The contribution from UPC events where both nuclei emit a bremsstrahlung photon is estimated using calculations from Ref.~\cite{Bertulani:1987tz} and is found to be negligible for photons with $|\eta|<2.4$ and $\et>3~\gev$.

\section*{\label{sec:Results}Results}

\begin{table}[t!]
\small
  \begin{center}
    \begin{tabular}{l|S[table-format=2.1]cS[table-format=1.1]S[table-format=2.1]|S[table-format=3.1]|c|S[table-format=3.0]}
\hline 
Selection			&  \text{\ggee} 	& CEP $gg\rightarrow\gamma\gamma$ 	&  \text{Hadronic} & \text{Other}  & \text{Total} & Signal  & \text{Data} 	\\ 
[-0.2em]
&  &  	&  \text{fakes} & \text{fakes} & \text{background} &  &  	\\

     \hline \hline
Preselection  					 	& 74 		& 4.7					&		6 & 	19  & 104 & 9.1  & 105 	\\

$N_\textrm{trk} = 0$ 						 	& 4.0 	& 4.5		&		6	& 19		&	33 & 8.7  & 39		\\

$\ptgg < 2$~\GeV{} 					 	& 3.5 	& 4.4				&	3		&	1.3 & 12.2 & 8.5  & 21  \\

Aco $<0.01$ 		 	& 1.3 	& 0.9				&		0.3	&	0.1  & 2.6 & 7.3  & 13 		\\
 \hline
Uncertainty				& 0.3		& 0.5		&		0.3					& 	0.1 & 0.7 & 1.5   & \\
\end{tabular}
\caption{The number of events accepted by the sequential selection requirements for data, compared to the number of background and signal events expected from the simulation. The signal simulation is based on calculations from Ref.~\cite{Klusek-Gawenda:2016euz}. In addition, the uncertainties on the expected number of events passing all selection requirements are given. }
\label{tab:Cutflow_opt}
\end{center}
\end{table}

\begin{figure}[t!]
\begin{center}
\includegraphics[width=.88\textwidth]{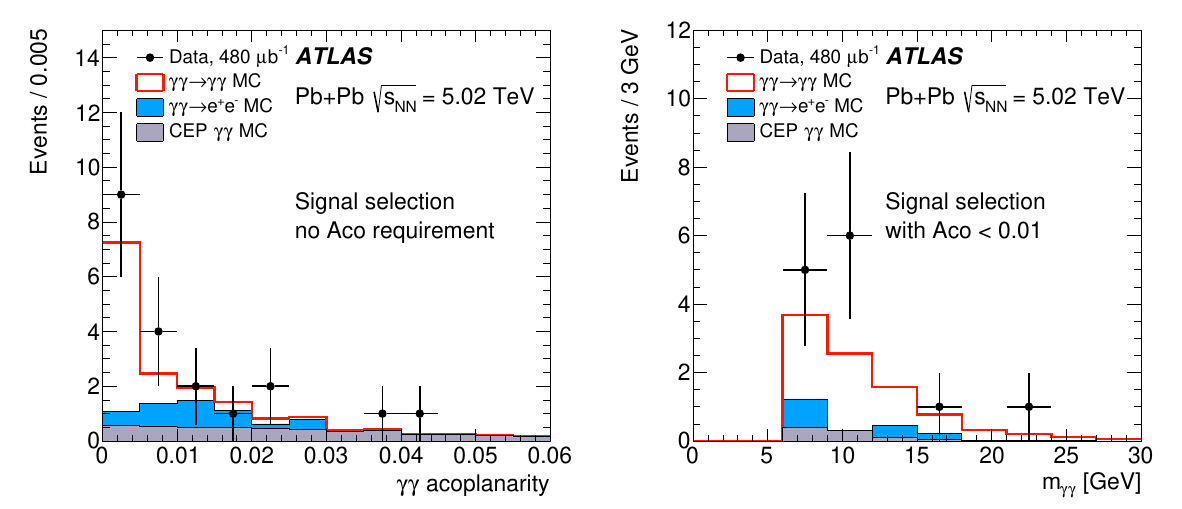}\put(-295,-8){{\footnotesize(a)}}\put(-95,-8){{\footnotesize(b)}}
\caption{Kinematic distributions for \gggg\ event candidates. (a) Diphoton acoplanarity before applying $\textrm{Aco}<0.01$ requirement. (b) Diphoton invariant mass after applying $\textrm{Aco}<0.01$ requirement. 
Data (points) are compared to MC predictions (histograms). The statistical uncertainties on the data are shown as vertical bars.}
\label{fig:LbL-control_opt}
\end{center}
\end{figure}

Photon kinematic distributions for events satisfying the selection criteria are shown in Figure~\ref{fig:LbL-control_opt}. 
The shape of the diphoton acoplanarity distribution for \ggee events in Figure~\ref{fig:LbL-control_opt}(a) reflects the trajectories of the electron and positron in the detector magnetic field, before they emit hard photons in their collisions with the ID material.
In total, 13 events are observed in data whereas 7.3 signal events and 2.6 background events are expected. 
In general, good agreement between data and MC simulation is observed. The effect of sequential selection requirements on the number of events selected is shown in Table~\ref{tab:Cutflow_opt}, for each of the data, signal and background samples.

To quantify an excess of events over the background expectation, a test statistic based on the profile likelihood ratio~\cite{Cowan:2010js} is used.
The $p$-value for the background-only hypothesis, defined as the probability for the background to fluctuate and give an excess of events as large or larger than that observed in the data, is found to be $5\times 10^{-6}$.
The $p$-value can be expressed in terms of Gaussian tail probabilities, which, given in units of standard deviation ($\sigma$), corresponds to a significance of $4.4\sigma$.
The expected $p$-value and significance (obtained before the fit of the signal-plus-background hypothesis to the data and using SM predictions from Ref.~\cite{Klusek-Gawenda:2016euz}) are $8\times 10^{-5}$ and $3.8\sigma$, respectively.

The cross section for the $\textrm{Pb+Pb}\,(\gamma\gamma)\rightarrow \textrm{Pb}^{(\ast)}\textrm{+}\textrm{Pb}^{(\ast)}\,\gamma\gamma$ process is measured in a fiducial phase space defined by the photon 
transverse energy $\et>3~\gev$, photon absolute pseudorapidity $|\eta|<2.4$, diphoton invariant mass greater than 6~\GeV, diphoton 
transverse momentum lower than 2~\GeV{} and diphoton acoplanarity below 0.01.
Experimentally, the fiducial cross section is given by

\begin{equation}
\label{EQN:CrossSectionFid}
\sigmafid = \frac{N_{\textrm{data}}-N_{\textrm{bkg}}}{C \times \int \! L \text{d} t}\,,
\end{equation}

where $N_{\textrm{data}}$ is the number of selected events in data, $N_{\textrm{bkg}}$ is the expected number of background events and $\int \! L \text{d} t$ is the integrated luminosity. 
The factor $C$ is used to correct for the net effect of the trigger efficiency, the diphoton reconstruction and PID efficiencies, as well as the impact of photon energy and angular resolution.
It is defined as the ratio of the number of generated signal events satisfying the selection criteria after particle reconstruction and detector simulation to the number of generated events satisfying the fiducial criteria before reconstruction.
The value of $C$ and its total uncertainty is determined to be $0.31 \pm 0.07$.
The dominant systematic uncertainties come from the uncertainties on the photon reconstruction and identification efficiencies. Other minor sources of uncertainty are the photon energy scale and resolution uncertainties and trigger efficiency uncertainty. 
In order to check for a potential model dependence, calculations from Ref.~\cite{Klusek-Gawenda:2016euz} are compared with predictions from Ref.~\cite{Enterria:2013yra}, and a negligible impact on the $C$-factor uncertainty is found.
Table~\ref{tab:AandCFactor} lists the separate contributions to the systematic uncertainty.
\begin{table}[t!]
\small
  \begin{center}
    \begin{tabular}{l|c}
\hline 
				Source of uncertainty			& Relative uncertainty 	\\
\hline \hline 
 Trigger 	& \ 5\%						\\
 Photon reco efficiency 	& 12\%	\\
Photon PID efficiency	& 16\%	\\	
 Photon energy scale 	& \ 7\%	\\ 
 Photon energy resolution 	& 11\%   \\
\hline 
 Total 				& 24\%						\\

\end{tabular}
\caption{Summary of systematic uncertainties. The table shows the relative systematic uncertainty on detector correction factor $C$ broken into its individual contributions. The total is obtained by adding them in quadrature.}
\label{tab:AandCFactor}
\end{center}
\end{table}
The uncertainty on the integrated luminosity is $6\%$.
It is derived following a methodology similar to that detailed in Refs.~\cite{Aad:2013ucp,Aaboud:2016hhf}, from a calibration of the luminosity scale using \textit{x}--\textit{y} beam-separation scans performed in December 2015.

The measured fiducial cross section is $\sigmafid = 70 \pm 24~\textrm{(stat.)} \pm 17~\textrm{(syst.)}$ nb, which is in agreement with the predicted values of $45 \pm 9$ nb~\cite{Enterria:2013yra} and $49 \pm 10$ nb~\cite{Klusek-Gawenda:2016euz} within uncertainties.

\section*{\label{sec:Conclusion}Conclusion}

In summary, this article presents evidence for the scattering of light by light in quasi-real photon interactions from 480 $\mu{\mathrm{b}}^{-1}$ of ultra-peripheral \PbPb\ collisions at \sqrtNN$=5.02~\TeV$ by the ATLAS experiment at the LHC. 
The statistical significance against the background-only hypothesis is found to be $4.4$ standard deviations.
After background subtraction and analysis corrections, the fiducial cross section for the $\textrm{Pb+Pb}\,(\gamma\gamma)\rightarrow \textrm{Pb}^{(\ast)}\textrm{+}\textrm{Pb}^{(\ast)}\,\gamma\gamma$ process 
was measured and is compatible with SM predictions.

The analysis is mostly limited by the amount of data available and the lower limit on transverse energy for reconstructed photons ($\et=3~\gev$), below which more signal is expected. Advancements on these two points would also allow for reconstruction of low-mass mesons decaying into two photons, which in turn could be used to improve detector calibration. The heavy-ion data yield is expected to double at the end of 2018 (and again increase tenfold after LHC Run 4, scheduled to start in 2026), which would significantly reduce the statistical uncertainty. Future upgrades of ATLAS, such as extended tracking acceptance from $|\eta| < 2.5$ to $|\eta| < 4.0$, will further improve this.


\section*{Acknowledgements}

We thank CERN for the very successful operation of the LHC, as well as the
support staff from our institutions without whom ATLAS could not be
operated efficiently.

We acknowledge the support of ANPCyT, Argentina; YerPhI, Armenia; ARC, Australia; BMWFW and FWF, Austria; ANAS, Azerbaijan; SSTC, Belarus; CNPq and FAPESP, Brazil; NSERC, NRC and CFI, Canada; CERN; CONICYT, Chile; CAS, MOST and NSFC, China; COLCIENCIAS, Colombia; MSMT CR, MPO CR and VSC CR, Czech Republic; DNRF and DNSRC, Denmark; IN2P3-CNRS, CEA-DRF/IRFU, France; SRNSFG, Georgia; BMBF, HGF, and MPG, Germany; GSRT, Greece; RGC, Hong Kong SAR, China; ISF, I-CORE and Benoziyo Center, Israel; INFN, Italy; MEXT and JSPS, Japan; CNRST, Morocco; NWO, Netherlands; RCN, Norway; MNiSW and NCN, Poland; FCT, Portugal; MNE/IFA, Romania; MES of Russia and NRC KI, Russian Federation; JINR; MESTD, Serbia; MSSR, Slovakia; ARRS and MIZ\v{S}, Slovenia; DST/NRF, South Africa; MINECO, Spain; SRC and Wallenberg Foundation, Sweden; SERI, SNSF and Cantons of Bern and Geneva, Switzerland; MOST, Taiwan; TAEK, Turkey; STFC, United Kingdom; DOE and NSF, United States of America. In addition, individual groups and members have received support from BCKDF, the Canada Council, CANARIE, CRC, Compute Canada, FQRNT, and the Ontario Innovation Trust, Canada; EPLANET, ERC, ERDF, FP7, Horizon 2020 and Marie Sk{\l}odowska-Curie Actions, European Union; Investissements d'Avenir Labex and Idex, ANR, R{\'e}gion Auvergne and Fondation Partager le Savoir, France; DFG and AvH Foundation, Germany; Herakleitos, Thales and Aristeia programmes co-financed by EU-ESF and the Greek NSRF; BSF, GIF and Minerva, Israel; BRF, Norway; CERCA Programme Generalitat de Catalunya, Generalitat Valenciana, Spain; the Royal Society and Leverhulme Trust, United Kingdom.

The crucial computing support from all WLCG partners is acknowledged gratefully, in particular from CERN, the ATLAS Tier-1 facilities at TRIUMF (Canada), NDGF (Denmark, Norway, Sweden), CC-IN2P3 (France), KIT/GridKA (Germany), INFN-CNAF (Italy), NL-T1 (Netherlands), PIC (Spain), ASGC (Taiwan), RAL (UK) and BNL (USA), the Tier-2 facilities worldwide and large non-WLCG resource providers. Major contributors of computing resources are listed in Ref.~\cite{ATL-GEN-PUB-2016-002}.





\section*{Supplementary information}

\subsection*{Exclusive dielectron production}

Exclusive dielectron pairs from the reaction $\textrm{Pb+Pb}\,(\gamma\gamma)\rightarrow \textrm{Pb}^{(\ast)}\textrm{+}\textrm{Pb}^{(\ast)}\,e^+e^-$ are used for various aspects of the nominal analysis, in particular to validate the EM calorimeter energy scale and resolution.

To select these \ggee candidates, events are required to pass the same trigger as in the diphoton selection.
Each electron is reconstructed from EM energy cluster in the calorimeter matched
to a track in the inner detector~\cite{Aad:2014fxa}. 
The electrons are required to have a transverse energy $\et>2.5~\GeV$ and pseudorapidity $|\eta|<2.47$ with the calorimeter transition region $1.37<|\eta|<1.52$ excluded.  
They are also required to meet “loose” identification criteria based on shower shape and track-quality variables~\cite{Aad:2014fxa}.
Candidate events are selected by requiring two oppositely charged electrons and no further charged-particle tracks coming from an interaction region.

Selected events are compared with Monte Carlo (MC) simulation based on the \starlight 1.1 model~\cite{Klein:2016yzr}, which combines the \PbPb\ equivalent photon approximation with the leading-order formula for \ggee.
The detector response is modelled using GEANT4~\cite{Agostinelli:2002hh,Aad:2010ah} and the simulated events are passed through the same reconstruction and analysis chain as the data.

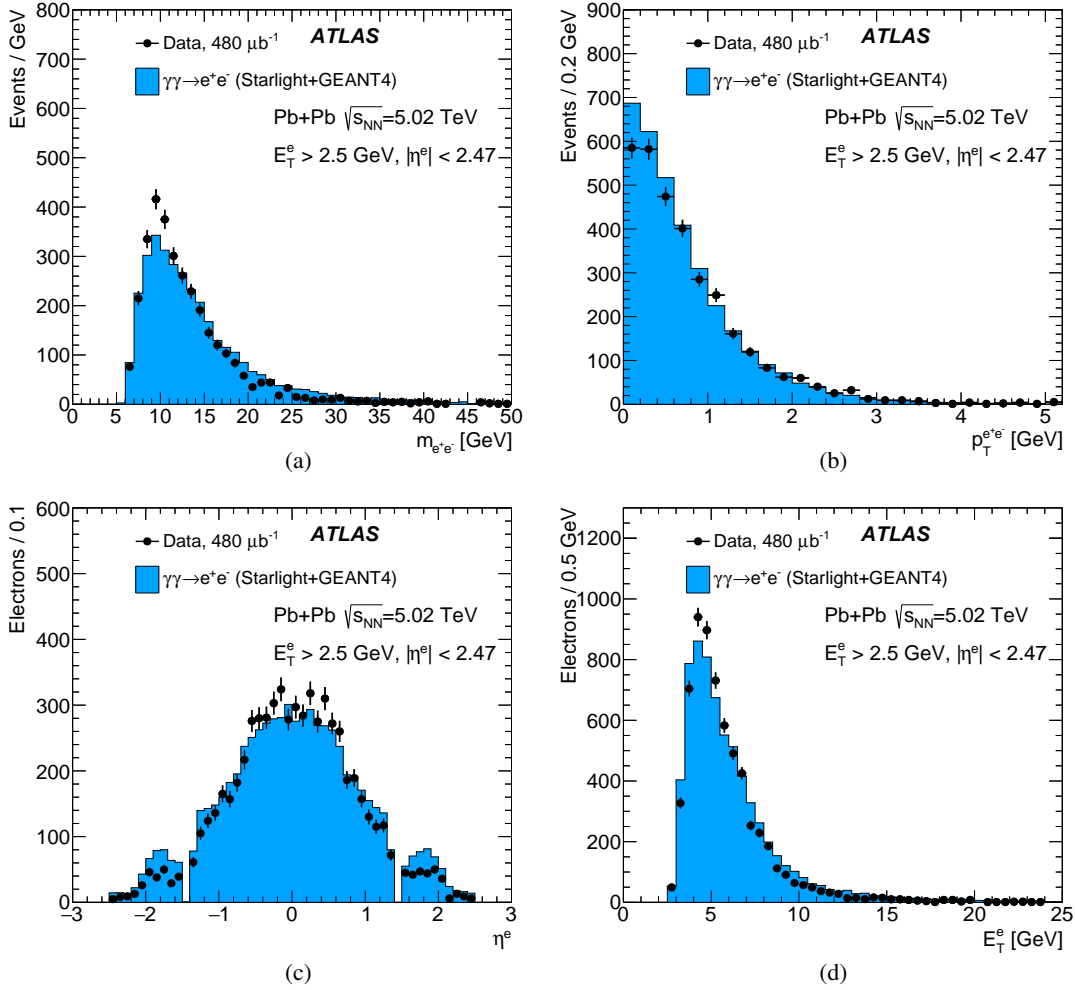
\begin{figure}[t!]
\begin{center}
\includegraphics[width=0.9\textwidth]{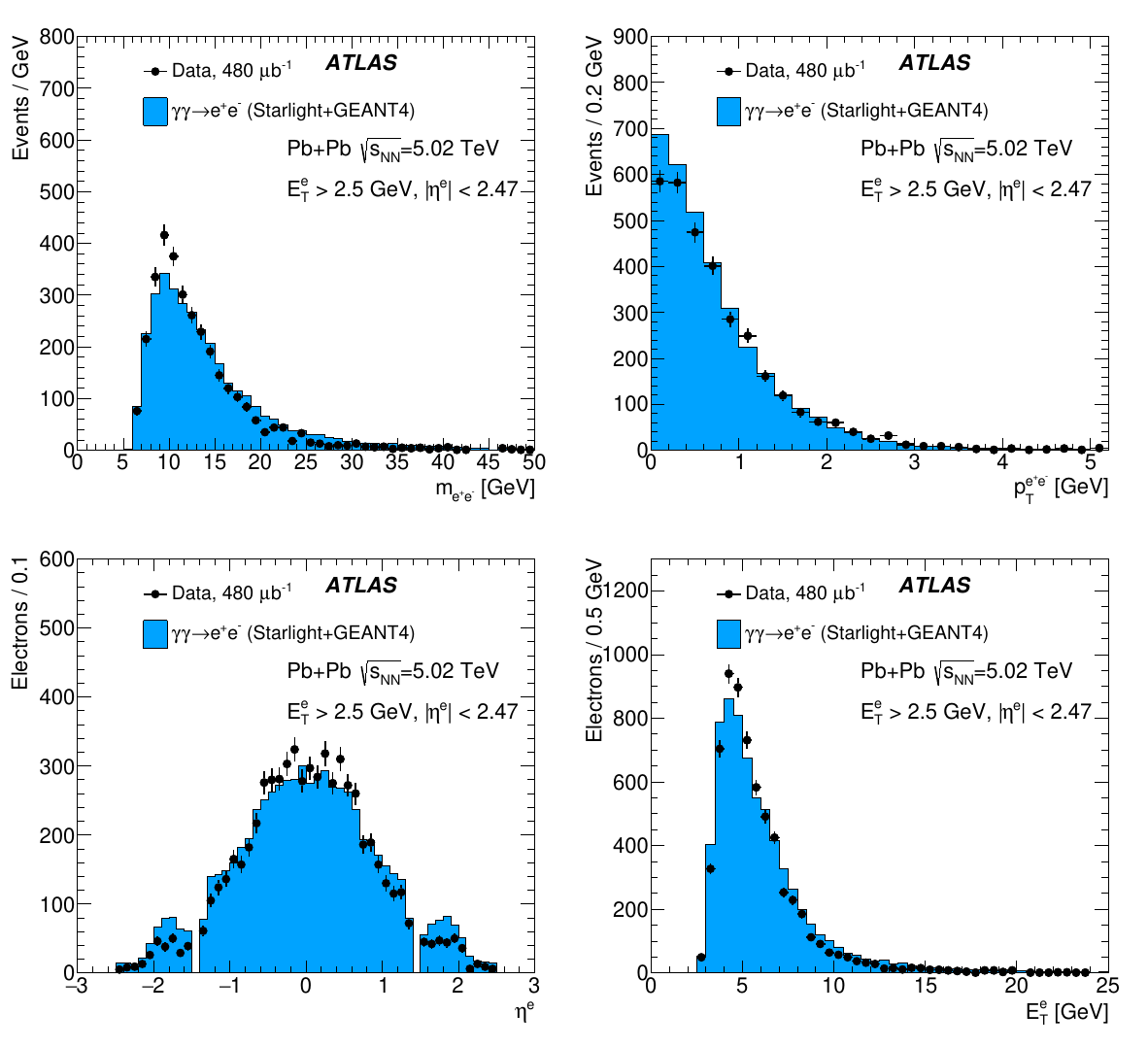}\put(-305,-3){{\footnotesize(c)}}\put(-105,-3){{\footnotesize(d)}}\put(-305,190){{\footnotesize(a)}}\put(-105,190){{\footnotesize(b)}}
\caption{Kinematic distributions for $\textrm{Pb+Pb}\,(\gamma\gamma)\rightarrow \textrm{Pb}^{(\ast)}\textrm{+}\textrm{Pb}^{(\ast)}\,e^+e^-$ event candidates: (a) dielectron mass, (b) dielectron \pt, (c) electron pseudorapidity and (d) electron transverse energy. Data (points) are compared to MC expectations (histograms). 
Electrons with $\et>2.5~\gev$ and $|\eta|<2.47$ excluding the calorimeter transition region $1.37<|\eta|<1.52$ are considered. 
 The statistical uncertainties on the data are shown as vertical bars. The uncertainty on the integrated luminosity, used to estimate the number of expected MC events, is 6\%.
}
\label{fig:ele-kin}
\end{center}
\end{figure}

Figure~\ref{fig:ele-kin} presents kinematic distributions of the dielectron system after the event selection.  
They show good agreement between the data and the QED prediction.
In total, 3216 dielectron events are selected in data and $3300\pm600$ events are expected from the simulation, where the uncertainty is due to limited knowledge of the initial photon fluxes. This modelling uncertainty is assigned as a global uncertainty and follows recommendations from Ref.~\cite{Enterria:2013yra}.

\subsection*{Validation of CEP  $gg\rightarrow\gamma\gamma$ background modelling}

\begin{figure}[t!]
\begin{center}
\includegraphics[width=0.43\textwidth]{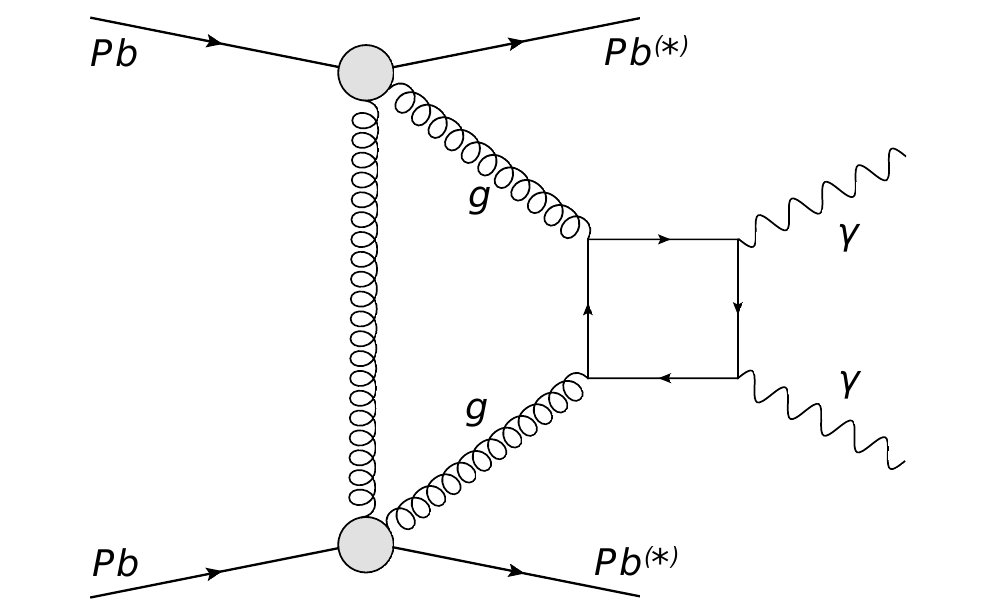}
\caption{Schematic diagram for the CEP $gg\rightarrow\gamma\gamma$ process production mechanism.}
\label{fig:cep}
\end{center}
\end{figure}

The central exclusive production (CEP) $gg\rightarrow\gamma\gamma$ is an important background process to consider in the nominal analysis, mainly due to similar two-photon final state and the ``peripheral'' nature of the interaction.
The CEP $gg\rightarrow\gamma\gamma$ occurs via the strong interaction through a quark loop in the exchange of two gluons in a colour-singlet state, which is schematically presented in Fig.~\ref{fig:cep}.

In Pb+Pb collisions this process can be modelled with \superchic~\cite{Harland-Lang:2015cta} MC generator, as suggested in Ref.~\cite{Enterria:2013yra}.
Since the exchanged objects are short-ranged comparing to the size of the Pb nucleus, the CEP occurs at relatively small impact parameters ($b$): typically twice the radius of the nuclei ($2 R$)~\cite{Baur:2001jj}.
Moreover, the exchanged objects would normally give a large momentum transfer to the nucleus~\cite{Baur:2001jj}, leading to moderate tails in the $\gamma\gamma$ acoplanarity (Aco).
These two effects would also result in a large probability of the outgoing ions to break-up (incoherent production) and in a strong suppression of the coherent CEP.
Due to small impact parameters in CEP, the coherent production is further altered by additional Coulomb excitations of the outgoing ions~\cite{Baur:2003ar}. 
The probability for the additional Coulomb break-up of at least one nucleus is estimated to be 80\% for $b=2 R$~\cite{Klein:2016yzr}.

When a nucleus breaks up, it produces neutrons at very small angles with respect to the Pb beams (forward neutrons).
They are measured in ATLAS using zero-degree calorimeters (ZDC), which are sensitive to neutrons and photons with $|\eta|>8.3$.
Therefore, to check the modelling of the CEP $gg\rightarrow\gamma\gamma$ background, an analysis of energy deposits in ZDC is performed.
The events are categorised for the signal ($\textrm{Aco}<0.01$) and the CEP-enhanced ($\textrm{Aco}>0.01$) regions.
To separate the ZDC signal from the noise of electronic modules, a calibrated ZDC energy greater than 40\% of the single neutron peak is required.

\begin{figure}[t!]
\begin{center}
\includegraphics[width=0.43\textwidth]{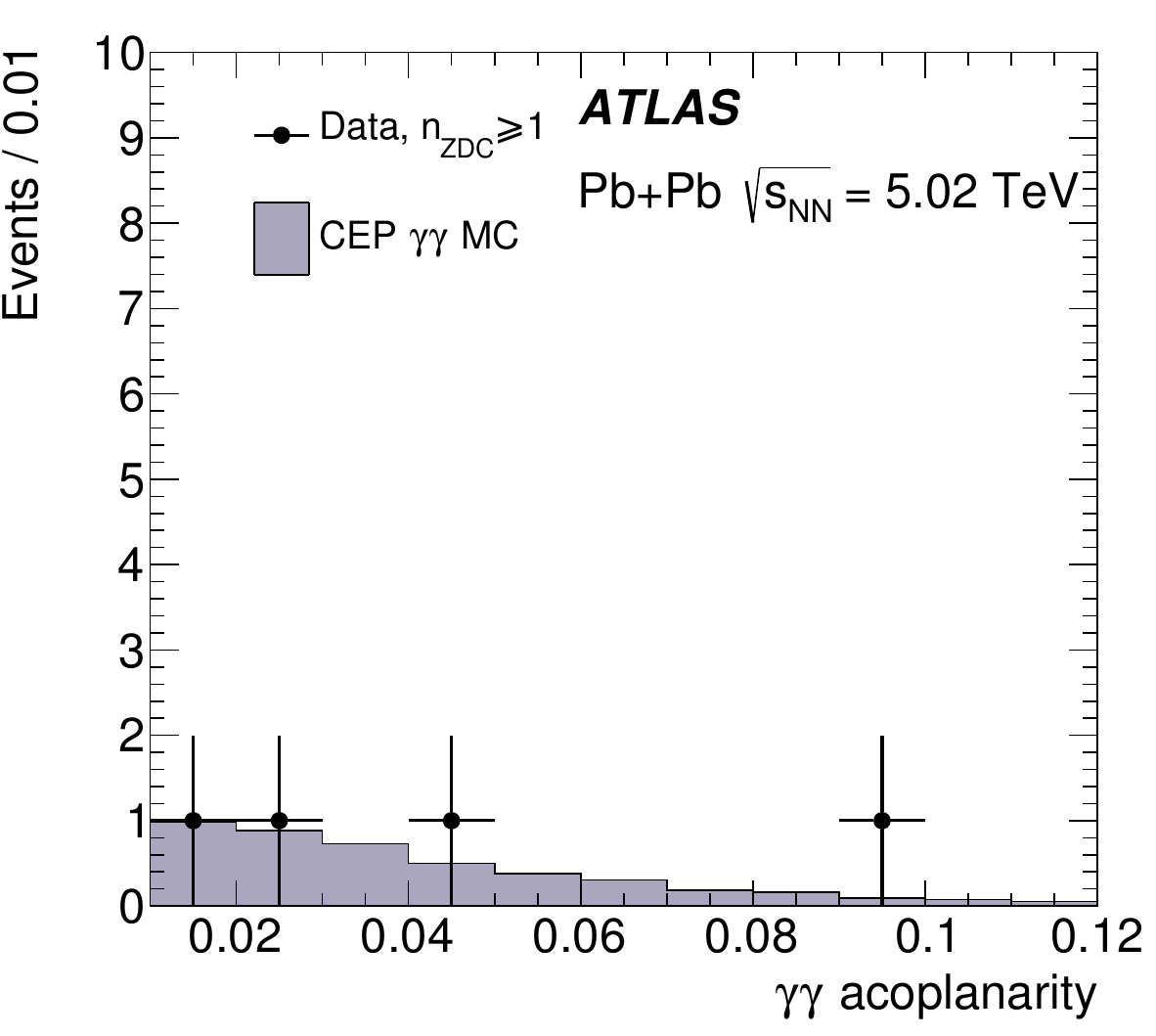}
\caption{Diphoton acoplanarity distribution observed in data (points) for events in CEP-enhanced region ($\textrm{Aco}>0.01$) with energy deposit in ZDC corresponding to multiple forward neutron emission. 
For comparison, CEP $gg\rightarrow\gamma\gamma$ MC predictions are also shown.
The statistical uncertainties on the data are presented as vertical bars.
}
\label{fig:aco_zdc}
\end{center}
\end{figure}

In the CEP-enhanced region, 4 events with no ZDC signal and
4 events with ZDC signal corresponding to multiple neutron emission (8 events in total) are observed in data, where 3.5 CEP $gg\rightarrow\gamma\gamma$ events are expected from the simulation.
A diphoton acoplanarity distribution for events with multiple forward neutron emission is presented in Fig.~\ref{fig:aco_zdc}, which tends to agree with the CEP $gg\rightarrow\gamma\gamma$ MC expectation.
This observation suggests that the transverse momentum transfer in incoherent heavy-ion CEP is comparable with the proton--proton case, which justifies the usage of \superchic~generator to model CEP $gg\rightarrow\gamma\gamma$ background contribution.

In the signal region, 
11 events with no ZDC signal and 
2 events with ZDC signal corresponding to exactly one neutron emission (13 events in total) are observed in data.
The expected event yield from CEP $gg\rightarrow\gamma\gamma$ MC is 0.9 events, however, events with one or more emitted neutrons are expected from the signal process, due to an excitation of the nuclear giant dipole resonance~\cite{Klein:2016yzr}. 

\newpage
\printbibliography


\newpage 
\section*{Author information}

\begin{flushleft}
{\Large The ATLAS Collaboration}

\bigskip

M.~Aaboud$^\textrm{\scriptsize 137d}$,
G.~Aad$^\textrm{\scriptsize 88}$,
B.~Abbott$^\textrm{\scriptsize 115}$,
J.~Abdallah$^\textrm{\scriptsize 8}$,
O.~Abdinov$^\textrm{\scriptsize 12}$$^{,*}$,
B.~Abeloos$^\textrm{\scriptsize 119}$,
S.H.~Abidi$^\textrm{\scriptsize 161}$,
O.S.~AbouZeid$^\textrm{\scriptsize 139}$,
N.L.~Abraham$^\textrm{\scriptsize 151}$,
H.~Abramowicz$^\textrm{\scriptsize 155}$
% ...
\bigskip
\\
% ...
$^{8}$ Department of Physics, The University of Texas at Arlington, Arlington TX, United States of America\\
% ...
$^{12}$ Institute of Physics, Azerbaijan Academy of Sciences, Baku, Azerbaijan\\
% ...
$^{88}$ CPPM, Aix-Marseille Universit{\'e} and CNRS/IN2P3, Marseille, France\\
% ...
$^{115}$ Homer L. Dodge Department of Physics and Astronomy, University of Oklahoma, Norman OK, United States of America\\
% ...
$^{119}$ LAL, Univ. Paris-Sud, CNRS/IN2P3, Universit{\'e} Paris-Saclay, Orsay, France\\
% ...
$^{137}$ $^{(a)}$ Facult{\'e} des Sciences Ain Chock, R{\'e}seau Universitaire de Physique des Hautes Energies - Universit{\'e} Hassan II, Casablanca; $^{(b)}$ Centre National de l'Energie des Sciences Techniques Nucleaires, Rabat; $^{(c)}$ Facult{\'e} des Sciences Semlalia, Universit{\'e} Cadi Ayyad, LPHEA-Marrakech; $^{(d)}$ Facult{\'e} des Sciences, Universit{\'e} Mohamed Premier and LPTPM, Oujda; $^{(e)}$ Facult{\'e} des sciences, Universit{\'e} Mohammed V, Rabat, Morocco\\
% ...
$^{139}$ Santa Cruz Institute for Particle Physics, University of California Santa Cruz, Santa Cruz CA, United States of America\\
% ...
$^{151}$ Department of Physics and Astronomy, University of Sussex, Brighton, United Kingdom\\
% ...
$^{155}$ Raymond and Beverly Sackler School of Physics and Astronomy, Tel Aviv University, Tel Aviv, Israel\\
% ...
$^{161}$ Department of Physics, University of Toronto, Toronto ON, Canada\\
% ...
$^{a}$ Also at Department of Physics, King's College London, London, United Kingdom\\
$^{b}$ Also at Institute of Physics, Azerbaijan Academy of Sciences, Baku, Azerbaijan\\
$^{c}$ Also at Novosibirsk State University, Novosibirsk, Russia\\
$^{d}$ Also at TRIUMF, Vancouver BC, Canada\\
$^{e}$ Also at Department of Physics {\&} Astronomy, University of Louisville, Louisville, KY, United States of America\\
% ...
$^{*}$ Deceased
\end{flushleft}

%

\end{document}